\documentstyle[12pt]{article}

\def\be{\begin{equation}}
\def\ee{\end{equation}}
\def\bea{\begin{eqnarray}}
\def\eea{\end{eqnarray}}
\def\ba{\begin{array}}
\def\ea{\end{array}}
\def\ben{\begin{enumerate}}
\def\een{\end{enumerate}}

\def\nnu{\nonumber}
\def\lll{\label}
 
\begin{document}
\newcommand{\half}{{\textstyle\frac{1}{2}}}
\newcommand{\eqn}[1]{(\ref{#1})}
\newcommand{\npb}[3]{ {\bf Nucl. Phys. B}{#1} ({#2}) {#3}}
\newcommand{\pr}[3]{ {\bf Phys. Rep. }{#1} ({#2}) {#3}}
\newcommand{\prl}[3]{ {\bf Phys. Rev. Lett. }{#1} ({#2}) {#3}}
\newcommand{\plb}[3]{ {\bf Phys. Lett. B}{#1} ({#2}) {#3}}
\newcommand{\prd}[3]{ {\bf Phys. Rev. D}{#1} ({#2}) {#3}}
\newcommand{\hepth}[1]{ [{\bf hep-th}/{#1}]}
\newcommand{\grqc}[1]{ [{\bf gr-qc}/{#1}]}
 
\def\a{\alpha}
\def\b{\beta}
\def\g{\gamma}\def\G{\Gamma}
\def\d{\delta}\def\D{\Delta}
\def\ep{\epsilon}
\def\et{\eta}
\def\z{\zeta}
\def\t{\theta}\def\T{\Theta}
\def\l{\lambda}\def\L{\Lambda}
\def\m{\mu}
\def\f{\phi}\def\F{\Phi}
\def\n{\nu}
\def\p{\psi}\def\P{\Psi}
\def\r{\rho}
\def\s{\sigma}\def\S{\Sigma}
\def\ta{\tau}
\def\x{\chi}
\def\o{\omega}\def\O{\Omega}
\def\k{\kappa}
\def\pa {\partial}
\def\ov{\over}
\def\br{\\}
\def\ud{\underline}
\begin{flushright}
UM-TH-00-11\\
SINP/TNP/00-15\\
hep-th/0006193\\
\end{flushright}
\bigskip\bigskip
\begin{center}
{\large\bf 
((F, D1), D3) Bound State, S-Duality and \\
Noncommutative Open String/Yang-Mills Theory}
\vskip1cm
{\sc J. X. Lu$^a$, 
Shibaji Roy$^b$ and
Harvendra Singh$^b$}
\vskip0.5cm
$^a$ Randall Physics Laboratory \\ University of Michigan, Ann Arbor, 
MI 48109-1120, USA
 \vskip 0.5cm 
$^b$ Theory Division, Saha Institute of Nuclear Physics,\\
1/AF Bidhannagar, Calcutta-700 064, India 
\vskip 0.5cm
E-mails: jxlu@umich.edu; roy, hsingh@tnp.saha.ernet.in
\end{center}
\bigskip
\centerline{\bf ABSTRACT}
\bigskip

\begin{quote}
  
We study decoupling limits and S-dualities for noncommutative open 
string/ Yang-Mills theory in a gravity setup by considering an $SL(2,Z)$ 
invariant supergravity solutions of the form ((F, D1), D3) bound state of 
type IIB string theory.  This configuration can be regarded as D3-brane 
solution with both electric and magnetic fields turned on along one of 
the spatial directions of the brane and preserves half of the space-time 
supersymmetries of the string theory. Our study indicates that there
exists a decoupling limit for which the resulting theory is an open
string theory defined in a geometry with noncommutativity in both 
space-time and space-space directions. We study S-duality of
this noncommutative open string (NCOS) and find that the same decoupling
limit in the S-dual description gives rise to a space-space
noncommutative Yang-Mills theory (NCYM). We also discuss independently 
the decoupling limit for NCYM in this D3 brane background. Here we find 
that S-duality of NCYM theory does not always give a NCOS theory. 
Instead, it can give an ordinary Yang-Mills with singular metric and 
infinitely large coupling. We also find that the open string coupling 
relation between the two S-duality related theories is modified such that
S-duality of a strongly coupled open-string/Yang-Mills theory does not
necessarily give a weakly coupled theory. The relevant gravity dual
descriptions of NCOS/NCYM are also given.
 
\end{quote} 

\newpage
\section{Introduction}
Noncommutative geometry has appeared on various occasions in string theory
starting from Witten's original proposal as a framework for open string
field theory \cite{witt}. While this appearance might seem natural in the
matrix
theory formulation of string/M-theory \cite{bfss,ban,BS,CDS}, a more direct 
approach to extract
the noncommutative nature of the underlying space in the conventional open
string theory set up has been given in
\cite{how,douh,chuh,ardas,seiw}. In these cases, the
open string boundary condition in the presence of non-zero constant NSNS
$B$-field plays a crucial role. In general, it is difficult to define a
consistent quantum field theory in a non-commutative space-time because of the
loss of manifest Lorentz covariance and causality. However, the fact that
they can be embedded in string theory in a particular background
suggests that they might exist as consistent quantum theory at least in
some special cases. So, we ought to understand the nature of these
theories in order to understand quantum gravity at very high energies
where our notion of space-time changes drastically.

The appearance of noncommutative geometry particularly in space-space
directions has been elucidated by Seiberg and Witten \cite{seiw} in
general setup by
identifying a decoupling limit in which closed strings decouple from the
open strings ending on D$p$-branes in the presence of non-zero $B$-field
(with only spatial components). The resulting theory is a Yang-Mills
theory defined in noncommutative spaces. This
decoupling limit has been studied for a special system by Hashimoto and
Itzhaki \cite{hasi} and also by Maldacena and Russo \cite{malr}. 
They studied D3-brane
supergravity solution of type IIB string theory with magnetic fields along
one of the spatial directions of the brane. The supergravity solution in 
this case is
nothing but the (D1,D3) bound state solution of type IIB string theory,
found in \cite{rust,bremm},
where there are infinite number of D-strings along, say $x^1$ direction of
D3-brane (lying along, say $x^1,x^2,x^3$ directions). Under the decoupling
limit,  this (D1,D3) system provides the gravity dual description of
${\cal N}=4$ SUSY noncommutative Yang-Mills
theory (NCYM). The decoupling limit in this case corresponds to a purely
field theoretic limit where the spatial directions $x^2,x^3$ become
noncommutative. The S-duality of the gravity
dual description of NCYM has been  
 considered by two of the present authors sometime ago 
\footnote{Unfortunately, 
we failed to recognize its conncetion to the
noncommutative open string theory (NCOS). This same gravity dual description
in the context of NCOS has also been given recently in \cite{gopmms}.} 
in \cite{luroyone}, which, as we know now, provides  the gravity dual 
description to the newly discovered noncommutative open string theory
(NCOS) \cite{seist,gopmms}.

On the other hand, if one considers D3-brane with an electric field along
one of the spatial directions of the brane, the corresponding supergravity
solution is given by (F,D3) bound state \cite{rust,lur} and is S-dual 
to (D1,D3)
system.  In this case there are infinite number of fundamental strings
along, say, $x^1$ direction of D3-brane.  
It has been shown recently \cite{seist,gopmms} that
even though it is not possible to obtain a field theory decoupling
limit,
a careful examination in this
case reveals a decoupling limit in which the resulting theory is 
 a noncommutative open string theory (NCOS) \cite{seist,gopmms}. 
This is consistent since in this case the coordinates $x^0,x^1$ become 
noncommutative and a field theory with a 
noncommutative time coordinate cannot be unitary in general 
\cite{seisut,gomm,ahagm}.
Moreover, it is shown in \cite{gopmms} that the S-dual of NCYM in the case
with purely
magnetic field gives just NCOS theory. The S-dual descriptions  of NCOS
in six and fewer spacetime dimensions are recently proposed and analysed in
\cite{gop,klem,berbss}.
 The gravity dual descriptions of these NCOS with pure electric 
fields are given in \cite{har}. S-duality in noncommutative gauge theories
has been discussed in \cite{gars,nuos}. 

It is then natural to ask whether there exists a decoupling limit, when we
consider D3-branes with both electric and magnetic fields turned
on along the brane world-volume,  such
that the resulting theory decouples from the closed strings (or bulk
supergravity).
Also if such a decoupling limit exists, what kind of a theory does it
correspond to and what is its strong coupling dual?
We would like to address this issue based on a
gravity consideration in this paper. In order
to study the decoupling limit we consider the $SL(2,Z)$ invariant 
((F,D1),D3) bound state solution of type IIB string theory in two
versions related by S-duality \footnote{ 
This solution has been constructed in \cite{luroytwo}.  There are
infinite number of $(p,q)$-strings of Schwarz type along one of the 
spatial directions of D3-brane. Thus we note that the electric
field and the magnetic field are parallel to each other and the
solution preserves half of the spacetime supersymmetries of
string theory.}. It is not a priori clear whether  decoupling limits 
for this solution exist and whether the corresponding decoupled theory 
is an NCYM or NCOS. 
We will first show that for D3 branes with both electric and magnetic
fields, there exists a decoupling limit such that the corresponding theory
can be described by an open string with both space-space and space-time
noncommutativity.
Then we will show that in the S-dual description the same NCOS decoupling
limit always gives rise to NCYM with only
space-space noncommutativity. We therefore show, in the present case,
 that the S-dual of NCOS theory is a NCYM theory, which is the converse
to what has been shown in \cite{gopmms} for purely magnetic field case.
 Next we will describe that there also exists an independent 
decoupling limit, using this D3 brane configuration, for NCYM
 with space-space noncommutativity. However,  we will show
that this same decoupling limit in the S-dual description of NCYM 
does not always give rise to an NCOS, unlike the purely magnetic field 
case. For example,
in some cases we end up with an ordinary Yang-Mills theory with a
singular metric and an infinitely large gauge coupling. This is because 
the magnetic field in the dual description (or the electric field in the 
original description) modifies the would-be critical elecric field to 
a non-critical one eventhough with the electric field in the original 
description we have had a decoupling limit for nicely behaved NCYM. 
Let us clarify this point further to see how the
magnetic field in the dual description controls the S-dual behaviour of
NCYM. The ratio of electric field to the 
magnetic field in this case is of the order $ 1/\a'^{1 + \b}$ in the
decoupling limit as
we will show later, where $\a'$ is the fundamental string constant
and the index $\beta \ge0$.  
For $ 0 \leq \b < 1$, the S-dual of NCYM is an 
ill-defined ordinary YM theory with a singular metric and an infinitely large 
gauge coupling. For $\b = 1$, we
end up with NCOS as the S-dual of NCYM with noncommutativity in both 
space-space and space-time directions. 
For $\b > 1$, the S-dual of NCYM is
also a NCOS but with only space-time noncommutativity. In other words, for 
$\b > 1$, the effect of magnetic field on this theory becomes
less important. For $\b > 1$, the parameters used to define the NCOS
or its gravity dual are independent of the actual value of $\b$. This
indicates that different scaling limits corresponding to different $\b$
with $\b > 1$ are actually equivalent, i.e., giving the same
NCOS\footnote{We thank the referee for pointing this out to us.}.
For $\b = 1$, the magnetic field is just right to have an effect on the
NCOS such that it causes the space-space noncommutativity and 
modifies the
relation between the open string coupling and its S-dual. 
For $0 \leq \b < 1$,
the magnetic field is too strong such that we do not even end up with a
well 
defined theory. In particular, for $\b = 0$, the electric field does not
reach its critical value. So $\b$ plays the role of an order parameter. The 
critical value is $\b = 1$. So NCYM and NCOS are related to
each other by S-duality when $\b \geq 1$. For $\b > 1$,
the coupling constants are related inversely to each other, therefore a 
strongly coupled theory is related to a weakly coupled theory by
S-duality. For $\b =
1$, this relation is modified and a strongly coupled theory does not
necessarily give rise to a weakly coupled theory by S-duality.   

	The organization of this paper is as follows. 
In section 2, we present the
general setup for the quantities relevant for the decoupling limit
in a flat background. We can take these quantities as the asymptotic
values of various fields in a gravity configuration. In section
3, we first show how to obtain the two versions related by S-duality for
the gravity configuration of ((F, D1), D3) bound state
\cite{luroytwo}. We then present these two versions explicitly for this
bound state.  In section 4, 
we first discuss the NCOS decoupling
limit with space-space and
space-time noncommutativity based on one version of the gravity
description of ((F, D1), D3) bound state. We find 
that the corresponding decoupling limit in the S-dual version always gives 
NCYM. Therefore, we have the NCYM as the S-dual of the NCOS. We then
discuss the decoupling limit for NCYM and study its S-duality. We find
that the S-duality of NCYM does not necessarily give NCOS.  
In section 5, we present both the
gravity dual description of the NCYM and that of NCOS in the respective 
decoupling limits. The conclusions drawn here agree completely with what
we obtain in section 4. In section 6 we clarify some confusions
which
might arise in sections 4 and 5 and  discuss possible quantizations of
the open string coupling.
   
While preparing this manuscript we became aware of a paper \cite{chew}
in which some related aspectes were also discussed.

\section{The General Set-up}

We are considering D3-brane to lie along $x^1,x^2,x^3$ spatial directions
and both electric and magnetic fields are turned
on along the $x^1$ direction.  
Since the field strengths $F_{01}~(F_{23})$ corresponding
to electric (magnetic) fields can be suitably traded for NSNS
$B$-field components $B_{01}~(B_{23})$, so D3-brane can be regarded as
living in a constant $g_{\m\n}$ and $B_{\m\n}$ background, where
$\m,\n=0,1,2,3$. The consideration here simplifies the discussion of the
decoupling limits presented in the following section.
The closed string metric we consider has the form

\be 
-g_{00}=g_{11}=g_1,~~~{\rm and}~~~g_{22}=g_{33}=g_2
\lll{1}
\ee
and the NS-NS field has the form
\be
B_{01}=-B_{10}=b_1,~~~{\rm and}~~~B_{23}=-B_{32}=b_2.
\lll{2}\ee
where $g_1,g_2,b_1,b_2$ are some constant parameters. 

Because of the presence of $B_{\m\n}$ field, the open string boundary
condition is not of purely Neumann type, but rather is a Neumann-Dirichlet
mixed  boundary condition \cite{acnj} given as:
\be
g_{\m\n}\pa_\sigma x^\n +2\pi\a' B_{\m\n}\pa_\tau x^\n|_{\rm boundary}=0
\lll{bound}
\ee
where $\pa_\s$ and $\pa_\tau$ are respectively the normal and tangential
derivatives to the world-sheet boundary.
We stress that the coordinates $x^\mu$ defined in the
above equation are the scaled ones used later in the paper.

The effective Seiberg-Witten  open string metric is given by

\be 
G_{\m\n}=g_{\m\n}-(2\pi\a')^2 (B~g^{-1}~B)_{\m\n}.\lll{4}\ee

Substituting $g_{\m\n}$ and $B_{\m\n}$ from eqs.\eqn{1} and  \eqn{2}
we obtain
\bea
&&-G_{00}= G_{11}\equiv G_1=g_1 -{(2\pi\a'b_1)^2\ov g_1}= g_1(1-\tilde
E^2)\lll{5}\br
&&G_{22}= G_{33}\equiv G_2=g_2 +{(2\pi\a'b_2)^2\ov g_2}= g_2(1+\tilde
B^2)
\lll{6}
\eea
where dimensionless electric field 
$\tilde E= E/E_{cr}= {2\pi\a'b_1\over g_1}$
with $E=b_1$ and the critical value of the electric field $E_{cr}=
g_1/2\pi\a'$.
Similarly we have defined dimensionless magnetic field
$\tilde B= B/B_0= {2\pi\a'b_2\over g_2}$
with $B=b_2$ and  $B_0 = g_2/2\pi\a'$.

The Seiberg-Witten relation for the antisymmetric non-commutativity
parameter is 

\be
\Theta^{\m\n}= 2\pi\a' \left( {1\ov g+ 2\pi\a' B}\right)^{\m\n}_A
\lll{7}\ee
where `A' denotes the antisymmetric part. We find from \eqn{1} and \eqn{2}
\bea
&&\Theta^{01}\equiv\Theta_1= {-(2\pi\a')^2 b_1\over(2\pi\a' b_1)^2-g_1^2}
={\tilde E\over E_{cr} (1-\tilde E^2)}\lll{8}\br 
&&\Theta^{23}\equiv\Theta_2= {(2\pi\a')^2 b_2\over(2\pi\a' b_2)^2+g_2^2}
={\tilde B\over B_0 (1+\tilde B^2)} \lll{9}
\eea
Also the open string coupling constant is given by

\be
G_s=g_s \left( { det G_{\m\n}\over
det(g_{\m\n}+2\pi\a'B_{\m\n})}\right)^{1/2}
\lll{10}\ee
where $g_s$ denotes the closed string coupling constant. From eqs.
\eqn{1},\eqn{2},\eqn{5} and \eqn{6} we obtain the open string coupling
of the form
\bea
G_s&=&g_s \bigg[1-{(2\pi\a'b_1)^2\ov
g_1^2}\bigg]^{1/2} \bigg[1+{(2\pi\a'b_2)^2\ov g_2^2}\bigg]^{1/2}\nnu 
\br &=& g_s (1-\tilde E^2)^{1/2}(1+\tilde B^2)^{1/2}. \lll{11}
\eea
It should be pointed out here that the dimensionless electric field
$\tilde E$ cannot exceed 1, otherwsie the open string coupling would
become imaginary. There is no such bound for the dimensionless magnetic
field $\tilde B$. When $|\tilde E|\to1$, the corresponding electric field
attains a critical value, i.e. $| E|\approx E_{cr}$ and then for a
finite value of $\tilde B,~g_s$ has to scale appropriately to make the
open string coupling $G_s$ finite. Similarly when $|\tilde E|<1$ and
$\tilde B\to\infty,~g_s$ has to scale accordingly to make $G_s$ finite.
But when both $|\tilde E|\to1$, and $\tilde B\to\infty$ $g_s$ can remain
finite if the two factors $(1-\tilde E^2)^{1/2}$ and $(1+\tilde B^2)^{1/2}$
compensate each other to make open string coupling in \eqn{11} finite.
We will point out what the resulting theory is in each of the cases in
the following sections. Unlike in purely electric case, a critical
electric field does not always give rise to a NCOS.

 Finally we note
that the effective open string metric \eqn{4} and the
noncommutativity parameter \eqn{7} has been obtained by looking at the
disk propagator at the boundary \cite{clnj,seiw}:
\be
\langle x^\m(\tau) x^\n(0)\rangle=-\a'G^{\m\n}{\rm ln}\tau^2 + i
{\Theta^{\m\n}\over2}\epsilon (\tau).
\lll{12}
\ee
We mention here that, in the decoupling limit, when the second term in 
the r.h.s. of \eqn{12} remains finite while the first term goes to zero 
as $\a'\to 0$, we get an NCYM theory as in
(D1,D3) (i.e. purely magnetic case). But when both the terms remain finite 
then we get NCOS theory
as in (F,D3) case. We will see in the
folowing sections how these two cases can be realised 
 in the decoupling limit of the same ((F,D1),D3) bound
state, i.e. D3 brane with both
electric and magnetic fields.

\section{((F,D1), D3) Bound State}

As we mentioned earlier 
((F,D1),D3) bound state solution of type IIB string theory corresponds
to D3-brane with both electric and magnetic field turned on along one of
the spatial directions of the brane. This solution has been explicitly
constructed in \cite{luroytwo} and is known to preserve half of the
spacetime supersymmetries of string theory. For the purpose of this
paper, we need to write this gravity configuration in two versions
related by S-duality. Since we have nonvanishing RR scalar $\chi$, the 
S-duality cannot be implemented in a simple fashion as is usually
done. The main reason is that the dilaton does not go to its
inverse under S-duality in the presence of $\chi$. Here we adopt the following
approach. We first write down the manifestly $SL(2,Z)$ covariant
((F,D1),D3) configuration in the Einstein frame for the metric, dilaton,
and the NSNS $B$-field in the NSNS sector and the RR scalar $\chi$, RR
 2-form $A_2$ and RR 4-form $A_4$ with its self-dual field strength 
$F_5$ in the RR sector as:
\bea
&& d S_E^2 = e^{U_0} (H H')^{1/4} [ H^{- 1} (- (dx^0)^2 + (d x^1)^2) + 
H'^{ - 1} ((d x^2)^2 + (d x^3)^2) + dr^2+r^2d\Omega_5^2],\nnu\br 
&&e^\phi = g_s \frac{H''}{\sqrt{H H'}},\nnu\br
&&\chi = \frac{p q (H - H') + g_s \chi_0 \,\Delta_{(p, q)} H'}{q^2 H +
g_s^2 (p - \chi_0 q)^2 H'},\nnu\br
&&2\pi \alpha' B = g_s e^{U_0} (p - \chi_0 q)\,\Delta^{- 1/2}_{(p, q, n)} 
H^{-1} d x^0\wedge d x^1 - \frac{q}{n} H'^{-1} d x^2 \wedge d
x^3,\nnu\br
&&A_2 = e^{U_0}\left[q g_s^{-1} - \chi_0 (p - \chi_0 q)\right] 
\Delta^{- 1/2}_{(p, q, n)} H^{-1} d x^0 \wedge d x^1 + \frac{p}{n}
d x^2 \wedge d x^3,\nnu\br
&&F_5 = 16 \pi n \a'^2 (\star \epsilon_5 + \epsilon_5)\lll{efc},
\eea
where harmonic functions $H, H'$ and $H''$ are defined as
\bea
&&H = 1 + Q_3/r^4,~~H' = 1 +  \frac{n^2 e^{2 U_0}} {\Delta_{(p, q, n)}}
\frac{Q_3}{ r^4},\nnu\br
&&H'' = 1 + \frac{g_s^{-1} q^2 + e^{2 U_0} n^2}{\Delta_{(p, q, n)}}
\frac{Q_3}{r^4}.\lll{ehf}
\eea
In the above, $p,q,n$ are integers\footnote{We can replace the
integral charges $p$ and $q$ in the above equations by the trigonometric
functions of two angles defined as:
\be
cos \theta = \frac{e^{U_0} n}{\sqrt{g_s^{-1} q^2 + e^{2 U_0} n^2}},~~
cos \a = \frac{\sqrt{g_s^{-1} q^2 + e^{2 U_0} n^2}} 
{\sqrt{g_s p^2 + g_s^{-1} q^2 + e^{2 U_0} n^2}}.
\ee We will make use of these definitions later.}. $n$ represents the number of 
$D3$-branes and is inert under S-duality (or in general under
$SL(2,Z)$). There
are actually an infinite number of $(p,q)$ strings of Schwarz type in
D3-brane world volume. To be precise, there is a single $(p,q)$ string per
$(2\pi)^2\a'$ area of the infinite $x^2,x^3$ plane of the D3-brane (note
that $(p,q)$ strings lie along $x^1$). $(p, q)$ transforms as a doublet
under $SL(2,Z)$. In particular, $p \to \hat p = q$ and $q\to \hat q = -
p$ under S-duality. Also,
$r=\sqrt{(x^4)^2+...+(x^9)^2},~d\Omega_5^2$ is the line element on unit 
5-sphere, $\star$ denotes the Hodge dual in 10-dimensions, $\epsilon_5$
is the volume form defined on the 5-sphere of unit radius,
 and $g_s$ is the closed string coupling constant and is given as $g_s =
e^{\phi_0}$ with $\phi_0$ the asymptotic value of the dilaton. 
$\chi_0$ is the asymptotic value for the RR scalar and we set it to zero 
for the rest of this paper.
This implies that for the asymptotic closed string
coupling, we still have $g_s \to 1/g_s$ under S-duality (but not for the
general coupling $e^\phi$). The constant $U_0$ is inert under $SL(2,Z)$.
This constant is chosen for the purpose of labeling different vacua of
non-perturbative type IIB string theory\footnote{The asymptotic region
of a gravity configuration corresponds to such a vaccum.}. For example,
$U_0 = 0$ corresponds to the supergravity vaccum while $U_0 = -
\phi_0/2$ corresponds to the vacuum for the perturbative type IIB string
theory. Usually, when we choose a special $U_0$ (or vacuum), we break
the $SL(2,Z)$ symmetry manifestly such as the above perturbative type
IIB string vaccum. With this $U_0$, the $Q_3, \Delta_{(p, q)}$ and $
\Delta_{(p, q, n)}$ are all $SL(2, Z)$ invariant and are given as
\be
\Delta_{(p, q)} = g_s (p - \chi_0 q)^2 + g_s^{- 1} q^2,~~
\Delta_{(p, q, n)} = \Delta_{(p, q)} + e^{2 U_0} n^2,~~
Q_3 = 4\pi \a'^2 \Delta_{(p, q, n)}^{1/2} e^{- 3 U_0}.\lll{esli}
\ee
Note that the string constant $\a'$ is also inert under $SL(2,Z)$ which
is consistent with the expression for the self-dual 5-form field
strength given in \eqn{efc}. With the above, one can check easily
that the harmonic functions $H$ and $H'$ are $SL(2,Z)$ invariant but
the harmonic function $H''$ is not. These are consistent with the 
Einstein-frame metric (which is $SL(2,Z)$ invariant) and the expression
for the dilaton. The transformation of the harmonic function $H''$ under
$SL(2,Z)$ is determined by that of the dilaton.

	With the above, we can easily write down  
the gravity configuration for the ((F, D1), D3) bound state in two
versions related by S-duality. They would look the same in form as 
given above in eq. \eqn{efc}. If we denote the above configuration as
Version A and denote the corresponding fields in the S-dual version
(called Version B) with a hat over the fields\footnote{We use $\hat A$ to
denote the field $A$ in the S-dual version.}, then the S-dual version
(i.e.,  Version B) is related to Version A through the following
relations (for vanishing $\chi_0$):
\be
\hat p = q,~~ \hat q = - p,~~ \hat g_s = \frac{1}{g_s}.\lll{abr}
\ee
Note that the constant $U_0$, the harmonic functions $H$ and $H'$ and
the 5-form field strength $F_5$ remain the same in the two versions
while the rest of the fields are related as
\be
e^{\hat \phi} = g_s^{-1} \frac{\hat H''}{\sqrt{H H'}},~~
\hat\chi = - \frac{pq (H - H')}{p^2 H + g_s^{- 2} q^2 H'},~~
2 \pi \a' \hat B = A_2,~~ \hat A_2 = - 2 \pi \a' B,\lll{abfr}
\ee
where the harmonic function $\hat H''$ is now given as
\be
\hat H'' = 1 + \frac{g_s p^2 + e^{2 U_0} n^2}{\Delta_{(p, q, n)}}
\frac{Q_3}{r^4}.\lll{sdh}
\ee

	Our purpose in this paper is to study decoupled theories and
their respective S-dualities in type IIB string theory. So we should use
the string-frame rather than the Einstein-frame description of the above
configuration. In Version A, we have the string-frame metric as
$d s^2 = e^{\phi/2} d s^2_E$ while in Version B, we have $d {\hat s}^2 =
e^{\hat \phi /2} d s^2_E$ with $d s_E^2$ the above Einstein-frame
metric. Given the asymptotic region of the gravity configuration
as a vacuum of the underlying string theory, we can choose, as is
usually done, either of the two S-duality related string-frame metric to have
the form $\eta_{MN}$ with $\eta_{MN} = (- 1,
1,\cdots, 1)$ ($M, N = 0, 1, \cdots, 9$) asymptotically. To be specific, 
we impose this
on Version A. This choice breaks the
$SL(2,Z)$ symmetry manifestly as is evident from $U_0 = - \phi_0/2$.
This can also be understood from the fact that
 we have made a preferable choice for strings over the other objects
such as 3-branes and 5-branes in this theory.

  With the above, the gravity configuration in Version A can be
   re-expressed as:

\bea
&&ds^2=
H''^{1/2}[H^{-1}(-(dx^0)^2+(dx^1)^2)+H'^{-1}((dx^2)^2+(dx^3)^2)+dr^2+r^2d\Omega_5^2 
], \nnu\br
&& e^{2\phi}= g_s^2 {H''^2\over H H'},~~ \chi = \frac{g_s^{-1} sin\theta\, tan\alpha\, (H - H')}{H sin^2\theta
+ H' tan^2\alpha},\nnu\br
&& 2\pi \alpha' B = sin\alpha\, H^{-1}dx^0\wedge dx^1 -
tan\theta H'^{-1}dx^2\wedge dx^3, \nnu\br
&& A_2 = g_s^{-1} sin\theta\, cos\alpha\, H^{ - 1} d x^0 \wedge d x^1 +
g_s^{- 1} tan\alpha\, cos^{- 1}\theta H'^{- 1} d x^2 \wedge d x^3, \nnu\br
&& F_5 = 16 \pi n \alpha'^2 (\star \epsilon_5 + \epsilon_5), \lll{17n}
\eea
where the harmonic functions are
\bea
&&H=1+ {4\pi g_s n \a'^2\over r^4} {1\over cos\theta cos\alpha},\nnu \br
&&H'=1+ {4\pi g_s n \a'^2\over r^4} { cos\theta cos\alpha },\nnu\br
&&H''=1+ {4\pi g_s n \a'^2\over r^4} { cos\alpha\over cos\theta},
\lll{17na} 
\eea
with
\bea
&& cos\theta={n\over\sqrt{q^2+n^2}},~~~
cos\alpha={\sqrt{q^2+n^2}\over\sqrt{(pg_s)^2+q^2+n^2}}.\lll{17nb}
\eea

The S-dual description (i.e., Version B) of the above  can be written 
as:
\bea
&&d {\hat s}^2 = {\hat g}_s {\hat H}''^{1/2}
[H^{-1}(-(dx^0)^2+(dx^1)^2)+H'^{-1}((dx^2)^2+(dx^3)^2)+dr^2+r^2d\Omega_5^2] 
\nnu\br
&&e^{2 \hat \phi} = {\hat g}_s^2 \frac{{\hat H}''^2}{H H'}\nnu\br
&&\hat \chi = - {\hat g}_s^{- 2} \frac{H''}{{\hat H}''} \chi \lll{18n}
\eea
where harmonic functions $H (= \hat H ), H' (= \hat H' )$ and 
$H''$ are the same as in Version A\footnote{Because of our special 
choice $U_0 = - \phi_0/2$, harmonic functions $H$ and $H'$ do not appear to be
manifestly invariant under S-duality.}, ${\hat g}_s = 1/g_s$ with $g_s$ the 
closed string coupling in the original version and the harmonic function 
${\hat H}''$ is not $SL(2, Z)$ invariant (neither is $H''$)  and is given as
\be
{\hat H}'' = 1 + \frac{4\pi n g_s \alpha'^2}{r^4} cos\alpha\,
cos\theta\,\left(1 + \frac{tan^2\alpha}{cos^2\theta}\right).
\lll{19n}\ee  
  
We also have $2\pi \alpha' {\hat B} = A_2$ and ${\hat A}_2 = - 2\pi
\alpha' B$ with $A_2$ and $B$ given in \eqn{17n}. The five-form remains 
the
same as in the original version. For later comparison, we give the
angles in the dual frame as,
\bea
&&cos \hat\theta = \frac{n}{\sqrt{g_s^2 p^2 + n^2}} = \left(1 +
\frac{tan^2 \a}{cos^2 \theta}\right)^{- 1/2},\nnu\br
&&cos \hat\a = \frac{\sqrt{g_s^2 p^2 + n^2}}{\sqrt{g_s^2 p^2 + q^2 +
n^2}} = \left(1 +
\frac{tan^2 \a}{cos^2 \theta}\right)^{ 1/2} cos\theta cos\a.\lll{ar}  
\eea

\section{The Decoupling Limit and S-Duality}

In this section we will be looking for decoupling limits such that the D3
brane in the presence of 
both electric and magnetic fields  decouples from the bulk closed
strings (or bulk gravity) and at the same time it gives rise to sensible
quantum theories for the decoupled D3 brane. This requires the open string 
coupling to remain fixed in the decoupling limit. In the following, we 
first discuss NCOS decoupling limit and study its S-duality (We will refer 
this as Case I).  Then we will discuss NCYM decoupling limit and its
S-duality (We call this as Case II).

\subsection{Case I: NCOS decoupling limit and S-duality}
Since we have given two 
versions of the gravity
configuration for the ((F, D1), D3) bound state related by S-duality
in the previous section, we should be 
able to show if there exists a connection 
between NCYM and NCOS by using purely the gravity description 
(if such decoupled theories  exist at all). In the case of purely
magnetic field on D3 brane it has been shown by Gopakumar et al
\cite{gopmms} 
that
starting from the known NCYM limit, one can use the S-duality and gauge 
transformation on the background $B$-field to get an NCOS theory. Thus 
in this case the S-dual of NCYM is a NCOS theory.
In the following, we will show that  in the presence of both electric
and magnetic fields, the decoupling limit for NCOS in Version A 
corresponds to a decoupling limit for NCYM in Version B. This implies
that the S-dual of NCOS is  NCYM even in the present case. 
 However, the converse 
is not quite true, i.e., the S-duality of NCYM does not always give an NCOS,
 as we will demonstrate later in subsection 4.2.    
  
	Let us start with the ((F,D1),D3) configuration in version A (eq.
\eqn{17n}). The fields discussed
in section 2 correspond to the asymptotic values of the
respective
fields in this gravity configuration. For general purpose, we assume
\be
x^{0, 1} = \sqrt{g_1} {\tilde x}^{0, 1}, ~~~~ x^{2, 3} = \sqrt{g_2} 
{\tilde x}^{2, 3},
\ee
where ${\tilde x}^\mu$ for $\mu = 0, 1, 2, 3$ remain fixed in the
decoupling limit. We then have
\be
{\tilde E} =  2\pi\a'b_1/g_1=sin\alpha,~~~~ 
{\tilde B} = 2\pi\a'b_2/g_2=- tan\theta,
\lll{20nn}
\ee
where $b_1 = g_1 sin\alpha/(2 \pi \alpha')$ and 
$b_2 = - g_2 tan\theta/(2\pi \alpha')$.
Using \eqn{5},\eqn{6},\eqn{8},\eqn{9} and \eqn{10}, we have
\bea
&&- G_{00} = G_{11} = G_1 = g_1 cos^2\alpha, ~~~~ G_{22} = G_{33} = G_2 = 
\frac{g_2}{cos^2\theta}\nnu\br
&& \Theta^{01} = \Theta_1 =  \frac{2\pi \alpha' sin\alpha}{g_1
cos^2\alpha},~~~~ \Theta^{23} = \Theta_2 = - \frac{2\pi\alpha' 
sin\theta cos\theta}{g_2}\nnu\br
&&G_s = g_s \frac{cos\alpha}{cos\theta}. \lll{20n}
\eea

In order to  have NCOS limit, we need 
at least the noncommutative parameter $\Theta_1$ 
in \eqn{20n} to be nonvanishing in the decoupling limit $\alpha' \rightarrow
0$, which requires 
$g_1 cos^2 \alpha \sim \alpha'$. So 
$ \alpha' G^{00} = - \alpha' G^{11}$ remain fixed and 
therefore the resulting theory
will be stringy rather than a field theory according to eq. \eqn{12} 
if the corresponding decoupling limit exists. Since $cos^2\alpha  \leq 1$,
we must
have $g_1 \sim \alpha'^\delta$ with $\delta \leq 1$. We limit ourselves to
$\delta < 1$ since we do not have decoupling for the special
case $\delta = 1$\footnote{For $\delta = 1$, $cos \alpha$ remains fixed as 
$\alpha' \rightarrow 0$. Given fixed open string coupling 
$G_s = g_s cos\alpha /cos\theta$, the string metric can only scale 
homogeneously in $\alpha'$ if $r \sim \sqrt{\alpha'}$, $g_2 \sim 
\alpha'$ and $cos \theta$ remains fixed. Then $g_s$ also remains fixed.
Since $\alpha' G^{\mu\nu}$ remain fixed for $\mu, \nu = 0, 1, 2, 3$ and
$\tilde E = sin \alpha$ is fixed and not equal to unity, 
we do not expect to have a decoupled open string theory. 
What is interesting in this case is
that both noncommutative parameters $\Theta_1$ and $\Theta_2$ are 
non-vanishing even as we take $\alpha'\rightarrow 0$. In other words,
in the near horizon region, the open string feels the geometry of the D3
brane to be noncommutative.}. Let us now examine the metric in Version A.
The
decoupling limit requires that the near-horizon region decouples
from the asymptotic flat region. In other words, we need the
metric describing the near-horizon region to scale homogeneously in 
certain power of $\alpha'$. This uniquely determines the scalings in
terms of $\alpha'$ for all the relevant quantities except for the 
$A_{01}$ component of the RR 2-form $A_2$ if $\delta < 1$. 
For the present purpose, we list the decoupling limit
collectively in the following:
\bea
&&r = \sqrt{{\tilde b}' \alpha'}\, u,~~ g_2 = \frac{\alpha'}{{\tilde
b}'},~~ g_1 = \left(\frac{\alpha'}{{\tilde b}'}\right)^\delta,\nnu\br
&& cos\theta = \frac{\tilde b}{{\tilde b}'},~~ cos^2\a =
\left(\frac{\alpha'}
{{\tilde b}'}\right)^{1 - \delta},\lll{21n}
\eea
where $u,~ {\tilde b}$ and ${\tilde b}'$ all remain fixed and $\delta <
1$. The above condition $cos\theta = {\tilde b}/{\tilde b'}$
is needed only for NCOS with noncommutativity in both space-space and
space-time directions. To include all the cases of NCOS, we should
 use
$sin\theta = c (\a'/\tilde b)^{\delta'}$ instead, with constant $c \leq 1$
and
parameter $\delta' \geq 0$. As will be shown in section 6, the $\delta'
\geq 0$ corresponds to $\b \geq 1$ which will be discussed in the
following subsection. In particular, $\delta' = 0$ corresponds to $\b =
1$. The $\delta' = 0$ is the same condition
as $cos \theta = \tilde b / \tilde b'$.  But for $\delta' > 0$, we have
the noncommutative parameter $\Theta_2$ vanishing and so we end up with
noncommutativity only in space-time directions, corresponding to vanishing
magnetic field.  Here we just concentrate on $\delta' = 0$ case.

If we calculate $G_1, G_2, \Theta_1, \Theta_2$ and $\tilde E$, using
eqs. \eqn{20nn}, \eqn{20n} and \eqn{21n}, we have
\bea
&&G_1 = \frac{\alpha'}{{\tilde b}'},~~~~G_2 = \frac{\alpha' {\tilde b}'}
{{\tilde b}^2},\nnu\br
&&\Theta_1 = 2 \pi {\tilde b}',~~~~ \Theta_2 = - 2\pi \frac{\tilde
b}{{\tilde b}'} \sqrt{{\tilde b}'^2 - {\tilde b}^2},\nnu\br
&& \tilde E = sin \alpha = 1 - \frac{1}{2} \left(\frac{\alpha'}{{\tilde
b}'}\right)^{1 - \delta}.\lll{22n}
\eea
Since $\Theta_1$ and $\Theta_2$ remain fixed (note that $\Theta_2$
vanishes in the special case of $\tilde b = \tilde b'$, corresponding to
vanishing magnetic field), we have
noncommutativity in both space-space and space-time directions. Further
we have $\tilde E \to
1$ as $\alpha' \to 0$ (since $\delta < 1$), i.e., electric field reaching
its critical value. We therefore expect a decoupled theory.
Since both $\alpha' G_1^{ - 1}$ and $ \alpha' G_2^{-1}$ remain fixed,
by looking at the correlator in eq.\eqn{12},
we must conclude that this decoupled theory is an open string theory
(rather than a field theory) 
defined on a spacetime with space-space and space-time noncommutative 
geometry. Therefore, in the decoupling limit \eqn{21n}, the ((F, D1), D3)
system
is described by a space-space and space-time noncommutative open string
theory.  Open strings are lying along $x^1$-direction, and 
$x^0, x^1$ coordinates are non-commuting (as $\Theta_1=$finite) as in
purely electric field case discussed by Seiberg et.al.\cite{seist} and 
Gopakumar et.al \cite{gopmms}. However, unlike in that case, 
we also have $x^2,x^3$ coordinates noncommuting (as $\Theta_2=$finite).
 As in purely electric case here also open 
strings cannot bend to form closed strings, i.e. closed strings
completely decouple. This can be understood as discussed by Seiberg et.al
\cite{seist},
from the expression of the electric field at critical value $E\approx
E_{cr}\sim \a'^{\delta-1} \to\infty$. So, it would require an infinite
amount of energy to bend such a string and therefore closed strings cannot
be formed. Another way to understand the decoupling of the open string
from the bulk closed strings is to compare the relevant energy scales.
The closed string scale is $M_s = 1/\sqrt{\a'}$ while the NCOS scale is
determined by the noncommutative parameter $\Theta_1$ as 
$M_{\rm eff} \sim 1/\sqrt{\tilde b'}$. This NCOS scale can be determined 
from: $1/(4\pi \alpha') \int \partial \tilde x^1 \partial \tilde
x^1 G_{11} = 1/(4\pi \tilde b') \int \partial \tilde x^1 \partial \tilde
x^1$. In other words, the effective $\a'_{\rm eff} = \tilde b'$.
In the limit $\a' \to 0$, the former becomes infinite while the
latter is fixed, therefore the open string decouples from the closed
strings. We can further understand this in the following way. Using  
$\a'_{\rm eff}$, the metric for NCOS can be scaled to  
$G_{\mu\nu} = \eta_{\mu\nu}$. The mass spectrum for NCOS is
\be
\a'_{\rm eff} M = \a'_{\rm eff} \left(p_0^2 - p_1^2 - p_2^2 - p_3^2\right) = N - 1.
\nnu   
\ee
where for simplicity we consider only bosonic string and $N$ is the
number of string excitations. Given that $\a'_{\rm eff}$ is fixed as 
$\a' \to 0$, the above equation is consistent with the nonzero $\a'
G^{-1}$ in the two-point function \eqn{12}. In other words, we have
undecoupled massive open string states from the massless ones 
and the energies of 
these finite exciations remain fixed in the decoupling limit. Let us see
if any of these open strings can be away from the brane and turn into
closed strings. If this is true, we should have the mass spectrum
for the closed strings (using the closed string metric) as
\be
\tilde b' \left(\frac{\a'}{\tilde b'}\right)^{1 - \delta} (p_0^2 -
p_1^2) - \tilde b' \left(p_2^2 + p_3^2\right) = 2 N + 2 \bar N - 4.
\nnu
\ee
The above equation cannot be satisfied unless the energy goes to
infinity as $\a' \to 0$ (since $\a'^{1 -\delta} \to 0$). We therefore
conclude that any NCOS with finite energy decouples from the closed
strings and is confined to the branes.

	 We like to point out that the NCOS discussed above 
 is insensitive to the
parameter $\delta$ even though the scaling limits are. This can be
understood by the fact that the NCOS is defined by the effective open
string metric, the noncommutative parameters and the open string
coupling, none of which has any dependence on the $\delta$ parameter.
This is also true for the gravity dual description of NCOS which will be
 discussed in the following section. Therefore,
the scaling limits for different $\delta < 1$ appears equivalent.

	Let us now examine what this NCOS would look like in the S-dual
theory, i.e., in Version B.
From the dual string metric and dual $B$-field, we have the following:
\bea
&&-{\hat G}_{00} = {\hat G}_{11} = {\hat G}_1 = g_1 g_s^{ -1} \left(1 -
sin^2\theta\,cos^2\alpha\right),\nnu\br
&&{\hat G}_{22} = {\hat G}_{33} = {\hat G}_2
= g_2 g_s^{- 1} \left(1 + \frac{tan^2\alpha}{cos^2\theta}\right),\nnu\br
&&{\hat \Theta}^{01} = {\hat \Theta}_1 = \frac{2\pi \alpha' sin\theta\,
cos\alpha}{ g_1 g_s^{ -1} \left(1 -
sin^2\theta\,cos^2\alpha\right)},~~ {\hat \Theta}^{23} = {\hat
\Theta}_2 = \frac{2\pi \alpha' tan\alpha/cos\theta}{g_2 g_s^{- 1} 
\left(1 + \frac{tan^2\alpha}{cos^2\theta}\right)},\nnu\br
&&{\hat G}_s = g_s^{-1} \left(1 - sin^2\theta\,cos^2\alpha\right)^{1/2}
\left(1 + \frac{tan^2\alpha}{cos^2\theta}\right)^{1/2},~~
\tilde {\hat E} = sin\theta\,cos\alpha,~~ \tilde{\hat B} =
\frac{tan\alpha}{cos\theta}. 
\lll{23n}
\eea
Using the decoupling limit in \eqn{21n}, we have 
\bea
&&{\hat G}_1 = G_s^{-1} \frac{\tilde b'}{\tilde b} \left(
\frac{\alpha'}{\tilde b'}\right)^{(1 + \delta)/2}, ~~{\hat G}_2 = G_s^{-
1} \left(\frac{\tilde b'}{\tilde b}\right)^3  \left(
\frac{\alpha'}{\tilde b'}\right)^{(1 + \delta)/2},\nnu\br
&& {\hat \Theta}_1 = 0,~~{\hat \Theta}_2 = 2\pi G_s 
\frac{{\tilde b}^2}{\tilde b'},~~{\hat G}_s = \frac{1}{G_s} 
\left(\frac{\tilde b'}{\tilde b}\right)^2,\nnu\br
&&{\tilde{\hat E}} \sim \alpha'^{(1 - \delta)/2},~~
\tilde{\hat B} \sim \alpha'^{- (1 - \delta)/2}.
\lll{24n}
\eea

As $\alpha' \rightarrow 0$, $\tilde {\hat E} \rightarrow 0$ and
$\tilde{\hat B} \rightarrow \infty$ ($\hat b_2$ fixed). Further we have 
$\alpha' {\hat G}_1^{- 1} \sim \alpha' {\hat G}_2^{- 1} \sim 
\alpha'^{(1 - \delta)/2} \rightarrow 0$, since $\delta < 1$, and
 the open string coupling ${\hat G}_s$ and noncommutative
parameter ${\hat \Theta}_2$ are all fixed. Therefore, we end up with NCYM as
expected. Notice that for 
$\delta = - 1$, ${\hat G}_1$ and ${\hat G}_2$ are also
fixed\footnote{For other values of $\delta < 1$, the metric for NCYM
appears to be singular while the coupling constant
remains fixed. Note that one cannot simply rescale the coordinates 
$\tilde x^\mu$ to make the metric nonsingular anymore because we 
assume the coordinates $\tilde x^\mu$ to be fixed from the outset.
Further, the two-point function \eqn{12} is defined with respect to
these fixed coordinates. If we rescale $\tilde x^\mu$ to $\bar x^\mu$ 
in order to have a nonsingular metric, we would end up with $< \bar x^\mu 
\bar x^\nu> \sim 0$, an ordinary field theory rather than a
noncommutative one. From \eqn{24n}, we see that the singular factor in
the metric is $(\a'/\tilde b')^{(1 + \delta)/2}$ which is also the
singular factor appearing in the asymptotic (i.e., $r \to \infty$) 
closed string metric in Version B. In other words, 
in the decoupling limit, unlike the usual case corresponding to 
$\delta = -1$, the asymptotic closed string metric is not  flat Minkowski
rather a flat Minkowski times the above singular factor 
$(\a'/\tilde b')^{(1 + \delta)/2}$. So, the singular 
behavior of the open string metric is being inherited from that of 
the asymptotic
closed string metric. As we know, the closed string is quantized 
perturbatively with respect to the flat Minkowski vacuum, which
usually corresponds to the asymptotic region of a gravity configuration.
In the present case, we could either choose $\delta = - 1$ with string
constant $\a'$ or we can have an effective description in the sense that
we have an effective string constant $\a'_{\rm eff} = \tilde b' 
(\a'/\tilde b')^{(1 - \delta)/2}$ (which can be obtained from 
$1/(4\pi \a') \int \partial x^M \partial x^N g_{MN} (r \to \infty) = 
1/[4 \pi \tilde b' (\a'/\tilde b')^{(1 - \delta)/2}] \int 
\partial y^M \partial y^N \eta_{MN}$, with $y^{0, 1} =  x^{0,
1}/\sqrt{g_1} = \tilde x^{0, 1},
y^{2, 3} = x^{2, 3}/\sqrt{g_1}, y^{4, \cdots, 9} = 
x^{4, \cdots, 9}/\sqrt{g_1}$), with again flat
Minkowski metric $\eta_{MN} = (- , +, \cdots, +)$. Using this effective
description, we calculate again the two-point function \eqn{12} (with 
respect to the same fixed $\tilde x^\mu$) and find 
that the new open string metric $ (\hat G_{\rm eff})_{\mu\nu}$ is 
nonsingular and is related to the original string metric as $\a'
\hat G^{\mu\nu} = \a'_{\rm eff} \hat G_{\rm eff}^{\mu\nu}$. Using the
effective open string metric $(\hat G_{\rm eff})_{\mu\nu}$, the NCYM is 
well-defined and is independent of the parameter $\delta < 1$. This is
true also for the gravity dual description which will be discussed in the
following section. In other words, many of these scaling limits 
corresponding to different $\delta < 1$ are actually equivalent.}.
The open string coupling relation given in \eqn{24n} gives 
$\hat G_s G_s = (\tilde b'/\tilde b)^2 > 1$, since  
$\tilde b'/\tilde b = 1/cos\theta > 1$. 
This implies that a strongly coupled theory does not necessarily give
rise to a weakly coupled theory after S-duality unless the
couplings for the strongly coupled theory is
greater than $(\tilde b'/\tilde b)^2$ in the presence of both electric 
and magnetic fields. This is
quite different from the purely electric or magnetic field case. The
reason for
this is actually simple. In the presence of both electric and magnetic
fields, we have the term $F\wedge F$ nonvanishing. In other words, we can 
effectively have an axion coupling. As we know, with nonvanishing axion,
 the coupling in the S-dual theory is not inversely related to the
coupling in the original theory. We will demonstrate this in the gravity
dual description by using the fact that the open string coupling is the
same as the closed string coupling at the IR and the fact that there
is an induced S-duality in the field theory
 from the S-duality of  type IIB string theory.
 In summary, we have
shown that the S-duality of NCOS in the presence of both electric and
magnetic fields gives an NCYM theory, like in the purely electric case  
\cite{gopmms}. Using the effective description discussed in the previous 
footnote, NCYM is essentially inert to the parameter $\delta < 1$. The
decoupling of the NCYM from the bulk gravity is the usual one and we do
not repeat the discussion here. 
In the following, we want to know whether the converse is true.

\subsection{Case II: NCYM decoupling limit and S-duality}

In this case we discuss the  decoupling limit for NCYM for the configuration
((F,D1),D3) in Version A (eq.\eqn{17n}).
To be specific, we insist\footnote{We could give a more general
discussion by insisting only $\alpha' G^{\mu\nu} \rightarrow 0$, i.e., 
for a field theory.} that $G_{\mu\nu}$ ($\mu,\nu = 0, 1, 2, 3$)
remain fixed (we choose them to be normalized to $\eta_{\mu\nu}$) 
as $\alpha' \rightarrow 0$. In order to have NCYM, we need 
$\Theta_2$ and $b_2 = -g_2\, tan \theta /(2\pi \alpha')$ to remain fixed.
From
\eqn{20n}, we have the following scalings:
\be
g_1 cos^2\alpha = 1, ~~ g_2 = \left(\frac{\alpha'}{\tilde b}\right)^2,
cos\theta = \frac{\alpha'}{\tilde b},\lll{25n}
\ee
where $\tilde b$ is a fixed parameter and is not related to the $\tilde b$ in 
the previous section. We also insist that the open
string coupling $G_s$ to be fixed which can be used to determine the
scaling behavior of the closed string coupling $g_s$. With 
$g_1 cos^2\alpha = 1$, one can check from \eqn{20n} that the
noncommutative
parameter $\Theta_1$ always vanishes as $\alpha' \rightarrow 0$. In the
decoupling limit, the gravity dual description of NCYM in the near horizon
region decouples from the asymptotic region. This requires that the 
near-horizon metric scales homogeneously in terms of certain power of
$\alpha'$. It is not difficult to check that for any $g_1$ satisfying
$g_1 cos^2\alpha = 1$, the scaling behavior for the radial coordinate
can be determined uniquely as
\be
r = \alpha' u,\lll{26n}
\ee
with $u$ fixed. We would like to point out one special case for which
the closed string coupling $g_s$ remains unscaled. From 
$G_s = g_s cos \alpha /cos \theta$, the choice $cos \alpha \sim \alpha'$
will do the job. 

Let us now examine the S-duality of the NCYM. 
For $g_1 cos^2\alpha = 1$, we have three cases which can be studied:
\begin{itemize}
\item
1) $g_1$ and $cos\alpha (\neq 1)$\footnote{The $cos \alpha = 1$
corresponds to vanishing electric field which is not our interest here. We
want to have both electric and magnetic field nonvanishing even though
one of them can be small. The case $cos \alpha = 0$ can never be
satisfied since we always assume nonvanishing integral charge $n$ for D3 
branes.} are both fixed and independent of the limit: $\alpha' 
\rightarrow 0$, 
\item
2) $g_1 = (\alpha'/\tilde b')^\delta$ with $\delta < 0$
and $cos \alpha = (\alpha'/\tilde b')^{-\delta/2}$
\item
 3) $g_1 \rightarrow
1$ and $sin \alpha = (\alpha'/\tilde b')^\b$ with $\b > 0$.
\end{itemize}
Note here that the parameters $\delta$ and $\tilde b'$ are different 
than those discussed in the previous case.    
{}From \eqn{23n}, we know that the quantity determining whether we have
NCOS in the dual theory or not is 
$\tilde{\hat E} = sin \theta cos\alpha$. Only
for $\tilde{\hat E} \rightarrow 1$ as $\alpha' \rightarrow 0$, we can
potentially have decoupled  NCOS. From the above decoupling limit for
NCYM, we have $sin\theta 
\rightarrow 1$. We therefore requires $cos\a \to 1$ for NCOS in the
S-dual description. So, cases 1) and 2) will certainly not give NCOS in the
S-dual theory.  Case 3) corresponds to the situation where the
electric field
times the closed string coupling is much smaller than the
magnetic field in
the original theory (Version A), i.e, $|b_1| g_s \ll |b_2|$. 
So it is clear now that except 
when one of the field is much weaker than the other
(or for purely electric or magnetic case)
the NCYM is not related to
NCOS by S-duality. Let us find out for each case above what is the
S-dual theory of NCYM.

\medskip

\noindent
Case 1): 

We exclude the special case $cos\alpha = 1$ which has been 
studied in \cite{gopmms}. We have the following scalings in the dual
theory,
\be
{\hat G}_1 = {\hat G}_2 \sim 1/\alpha',~~{\hat \Theta}_1 \sim \alpha'^2,
~~{\hat \Theta}_2 \sim \alpha',~~ {\hat G}_s \sim 1/\alpha'^2,~~{\hat b}_1 \sim
1/\alpha'^2,~~{\hat b}_2 \sim 1/\alpha'.
\lll{27n}\ee
{}From the above we see that we have a field theory (since 
$\alpha' {\hat G}_1^{-1} = \alpha' {\hat G}_2^{-1} \sim \alpha'^2 
\rightarrow 0$) according to \eqn{12} but defined in a commuting geometry
\footnote{Since we assume $\tilde x^{\mu}$ to be fixed, 
$< \tilde x \tilde x> \sim 0$ just as in a ordinary Yang-Mills case in 
the decoupling limit.  The
coupling still blows up. We do not know how to make sense of such a
theory.}.
But this theory is bad since it has an infinitely
large open string coupling even though the singular metric
can be brought to a nonsingular one as we discussed in 
footnote 9. Both the electric field ${\hat b}_1 \sim 1/
\alpha'^2$ (but $\tilde {\hat E} = cos\alpha$ fixed) and magnetic field
${\hat b}_2 = 1/\alpha'$ (the $\tilde {\hat B} = 1/\alpha'$ still large) are 
infinitely large.

\medskip

\noindent
Case 2): 

In this case we have the following scalings:
\be
{\hat G}_1 = {\hat G}_2 \sim \alpha'^{-1 +\delta/2},~~{\hat \Theta}_1 
\sim \alpha'^{2 - \delta}, ~~ {\hat \Theta}_2 \sim \alpha',
\ee
with ${\hat b}_1, {\hat b}_2$ and ${\hat G}_s$ remain the same as in
case 1). Recall
that we have $\delta < 0$, so this case is not much different from case 1)
as expected.

\medskip

\noindent
Case 3):  

This is the case where we expect to have NCOS. Let us list the 
relevant parameters,
\bea
&&\alpha'\rightarrow 0,
~~{\hat b}_1 = \frac{\tilde b}{2\pi G_s \alpha'^2}\left[
1 - \frac{1}{2}\left(\frac{\alpha'}{\tilde b}\right)^2\right],~~
{\hat b}_2 = \frac{1}{2\pi G_s \tilde b'} 
\left(\frac{\alpha'}{\tilde b'}\right)^{\b - 1},\nnu\br
&&\tilde{\hat E} = 1 - \frac{1}{2} 
\left(\frac{\alpha'}{\tilde b}\right)^2 
\left[1 + \left(\frac{\tilde b}{\tilde b'}\right)^2 
\left(\frac{\alpha'}{\tilde b'}\right)^{2(\b - 1)}\right],\nnu\br
&&
{\hat G}_1 = {\hat G}_2 = \frac{\alpha'}{G_s \tilde b} 
\left[1 + \left(\frac{\tilde b}{\tilde b'}\right)^2 
\left(\frac{\alpha'}{\tilde b'}\right)^{2(\b - 1)}\right],\nnu\br
&&{\hat \Theta}_1 = \frac{2\pi G_s \tilde b}{\left[1 + 
\left(\frac{\tilde b}{\tilde b'}\right)^2 
\left(\frac{\alpha'}{\tilde b'}\right)^{2(\b - 1)}\right]},~~
{\hat \Theta}_2 =\frac{2\pi G_s \tilde b}{\left[1 + 
\left(\frac{\tilde b}{\tilde b'}\right)^2 
\left(\frac{\alpha'}{\tilde b'}\right)^{2(\b -
1)}\right]}\frac{\tilde b}{\tilde b'} \left(\frac{\alpha'}{\tilde b'}
\right)^{\b - 1},\nnu\br
&&{\hat G}_s = \frac{ \left[1 + \left(\frac{\tilde b}{\tilde b'}\right)^2 
\left(\frac{\alpha'}{\tilde b'}\right)^{2(\b - 1)}\right]}{G_s}.
\lll{29n}
\eea
From above, we have basically three sub cases depending on the range of
the parameter $\b$. Remember that 
we always have  $\tilde {\hat E} \rightarrow 1$ as $\alpha'
\rightarrow 0$. For $\b \ge 1$, we have $\alpha' {\hat G}_1^{-1} = 
\alpha' {\hat G}_2^{- 1} = {\rm fixed}$, and  the open  string coupling 
${\hat G}_s$ and at least one of the noncommutative parameters are also
fixed. Therefore we end up
with a NCOS. However, the $\b > 1$ case differs from $\b = 1$
case in that the former has only noncommutativity along space-time directions
and the open string coupling is inversely related to its S-dual,
while the latter has noncommutativity not only in space-time directions
but also in space-space directions and the relation between the open
string couplings related by S-duality has been modified. We would like 
to stress that for $\b > 1$, the parameters such as the metric, 
noncommutative parameters and the open string coupling which define 
the NCOS theory are independent of the parameter $\b > 1$. The same is
true for the gravity dual description which will be discussed in the 
following section. This indicates that many of the scaling limits 
corresponding to different $\b > 1$ are actually equivalent. This $b >
1$ case corresponds to the $\delta' > 0$ case discussed in the previous 
sub-section. The precise relation between this $\b$ and the $\delta$ and 
$\delta'$ discussed in the previous sub-section will be given in section
6. For $\b = 1$, corresponding to $\delta' = 0$ discussed in the
previous sub-section, from the above 
equation \eqn{29n}, once again we have $\hat G_s G_s = 1 + (\tilde b/\tilde b')^2
> 1$. This implies that in the present case a weakly coupled theory
cannot be obtained from  a strongly coupled theory by S-duality unless
the strongly coupled theory has a coupling greater than $1 + (\tilde b/\tilde
b')^2$. 
  For $0 < \b < 1$, we end up
with an ill-behaved field theory  as in cases 1) and 2) discussed
above with the open string coupling blowing up even though we
now have reached the critical electric field limit\footnote
{We would
like to stress that the present discussion, i.e., NCYM and its S-duality
or Case II, is independent of what we have discussed in Case I. In particular, many
parameters used here are different, in definition, from those used in
Case I even though we often use the same symbols. For example, the
parameters $\tilde b$ and $\tilde b'$ here are different from those used
in Case I. Even the string constant $\a'$ may be different. But we
expect that the conclusions drawn under the same conditions should be 
the same, for example, we always have $\hat G_s G_s > 1$ in the presence
of both electric and magnetic fields. Another example is that in case I,
the limit $\tilde b' \to \tilde b$ gives pure electric case while here
the  corresponding limit is $\tilde b /\tilde b' \to$ zero where we
should take $\tilde b' \to \infty$ while keeping $\tilde b$  fixed.
These two cases can be identified if we exchange 
$cos\theta \leftrightarrow cos\a$ which can be recognized through the 
respective decoupling limits and the S-duality. Some relations between
the parameters in the two cases will be further clarified in section 6.}.
Unlike in cases 1) and 2), the singular metric here cannot be brought to
a non-singular one following the description given in footnote 9.

  The decoupling of NCOS for $\b \geq 1$ from the closed strings can be 
discussed similarly as we did in the previous subsection and we will not 
repeat it here.

	The above discussion indicates that the effect of one of the
fields on the other really drops out if $\b > 1$. This effect
becomes important for $\b = 1$, a sort of critical value. As 
$\b < 1$, the open string theory flows to an ill-defined
strongly-coupled field theory. Thus $\b$ plays the role of some kind of
an order parameter. 

	With this understanding of the relationship between NCYM and
NCOS, we will present the gravity dual descriptions for each case
discussed here in the following section.

\section{The Gravity Dual Descriptions}

	Here we present the gravity dual of the NCOS discussed first
(Case I) in the
previous section. The decoupling limit can be collectively given as,
\bea
&&\a'\to0,~~ u={r\over\sqrt{\tilde b'}\sqrt{\alpha'}}={\rm
fixed},~~~g_1 = \left(\frac{\alpha'}{\tilde b'}\right)^\delta,~~ g_2 = 
\frac{\alpha'}{\tilde b}\nnu,\br
&&cos\alpha=\left({\alpha'\over \tilde
b'}\right)^{(1 - \delta)/2},~~~cos\theta={\tilde b\over\tilde b'},\nnu\br
&& R^4= {\rm fixed}= 4\pi g_s n {cos\a\over cos\theta}= 4\pi G_s n,
\lll{18}\eea
where the parameter $\delta < 1$.
 The
closed string coupling can be obtained from the last expression in
\eqn{18} as,
\be
g_s={G_s\tilde b\over \tilde b'} \left(\frac{\tilde
b'}{\alpha'}\right)^{1 - \delta \over 2}.
\lll{19}\ee
Under this decoupling limit the harmonic functions in \eqn{17na} take the
following form;
\be
H={R^4 \over u^4 \tilde b'^2}\left(\frac{\tilde
b'}{\alpha'}\right)^{1 - \delta},~~H'=1+{R^4\tilde b^2\over u^4
\tilde b'^4},~~H''=1+{R^4\over u^4\tilde b'^2}.
\lll{20}
\ee
So the metric, dilaton, axion,  $B$-field and the RR 2-form  
 now reduce to 
\bea
&& ds^2 = \a' \bar{h}^{-1/2}\left\{{u^2\over R^2}\left[-(d\tilde x^0)^2
+ (d\tilde x^1)^2 + \bar{h}' ((d\tilde x^2)^2
+(d\tilde x^3)^2)\right] + R^2 \left[\frac{du^2}{u^2} +
d\Omega_5^2\right]\right\},\nnu\br
&& e^{2\phi}= G_s^2 {\tilde b^2\over\tilde
b'^2}{\bar{h}'\over\bar{h}^2}  \nnu\br
&& \chi = \frac{\sqrt{\tilde b'^2 - {\tilde b}^2}}{\tilde b  G_s}
{\bar h},\nnu\br
&&
2\pi \alpha' B = \a'\,{u^4\tilde b'\over R^4} d\tilde x^0\wedge d\tilde x^1 -
\a'\,{\sqrt{\tilde b'^2 -\tilde b^2}\over \tilde b} {u^4\tilde b'\over R^4}
\bar{h}' d\tilde x^2\wedge d\tilde x^3 ,\nnu\br
&& A_2 = {\a'}^2 \frac{u^4}{g_1 G_s R^4}
{\sqrt{\tilde b'^2 -\tilde b^2}\over \tilde b} d\tilde x^0\wedge 
d\tilde x^1 + \alpha' \left(\frac{\tilde b'}{\tilde b}\right)^3
\frac{\tilde b  u^4}{R^4 G_s} {\bar h'} d\tilde x^2\wedge d\tilde x^3,   
\lll{21}\eea
where we have defined $\bar{h}={1\over 1+a^4u^4},~\bar{h}'={1\over {\tilde b^2\over\tilde
b'^2} + a^4 u^4}$ with $a^4={\tilde b'^2\over R^4}$. As one can see that
$g_1$ appears only in the $01$-component of the RR two-form $A_2$,
indicating that the scaling behaviour of this component is dependent on
the parameter $\delta$. However, this component measured in terms of
$\a'$ is proportional to $\a'^{1 - \delta}$ which vanishes as $\a' \to
0$ since $1 - \delta > 0$. So the gravity dual description is
independent of the $\delta < 1$. 
 
Note from the form of the metric in \eqn{21} that as
 $u\to 0$, the metric reduces
to $AdS_5\times S^5$ form as
expected and it starts deviating from this form at $u\sim {R
\sqrt{\tilde b}\over {\tilde b'}}$.  We also notice from \eqn{21}
that as $u\to 0,~e^{2\phi}\to G_s^2$ and 
 so the open string coupling is the same as the closed string coupling 
at the IR. This occurs always as noticed in \cite{luroytwo,luroyone}.
Also the axion $\chi$ at the IR does not vanish but reaches a constant
value  $ (\tilde b'^2 - {\tilde b}^2)^{1/2}/(G_s \tilde
 b)$. The S-dual dilaton is related to the original one through
the relation $e^{\hat \phi} = e^\phi (\chi^2 + e^{- 2\phi})$. At the IR,
 we have $e^{\hat \phi} = \tilde b'^2 / (G_s\tilde b^2)$. But the closed
 string coupling $e^{\hat \phi}$ at the IR gives the open string
 coupling ${\hat G}_s$. So we provide an explanation for the S-dual
coupling relation, derived in the previous section, using the S-duality
of type IIB string theory. 
 This is another way to understand why
 in the presence of both electric and magnetic fields, a strongly
 coupled theory does not necessarily give a weakly coupled S-dual theory.
 It is easy to check that if $\tilde b = \tilde b'$, i.e., one
 of the field vanishes, then $\chi = 0$ at the IR and we recover
 the simple inverse relation for the couplings. Note, however, that at UV
the closed string coupling blows up.

	Now we move on to give the S-dual of the above gravity dual
description of NCOS, i.e., the gravity dual of NCYM. Using \eqn{18n},
\eqn{19n}, \eqn{18} and \eqn{20}, we have
\bea
&&d{\hat s}^2 = \alpha' \frac{\tilde b'}{G_s \tilde
b} \left\{{u^2\over R^2}\left[- (d\tilde x^0)^2
+ (d\tilde x^1)^2 + \bar{h}' ( (d\tilde x^2)^2
+ (d\tilde x^3)^2)\right] + R^2 \left[\frac{du^2}{u^2} +
d\Omega_5^2\right]\right\}, \nnu\br 
&&e^{ \hat \phi} = \frac{\tilde b'}{G_s \tilde b} \bar{h}'^{1/2},~~
{\hat \chi} = - \frac{G_s \tilde b}{\tilde b'} \frac{\sqrt{\tilde b'^2
- {\tilde b}^2}}{\tilde b'},\lll{22}
\eea
where $\bar h'$ is the same as that defined for NCOS. We also have  $ 2\pi
\alpha' {\hat B} = A_2$ and  
${\hat A}_2 = -  2\pi \alpha' B_2$ with $B_2$ and $A_2$ given in
\eqn{21}.
At the IR, i.e. for $u \rightarrow 0$, the metric describes $AdS_5 \times
S^5$ as expected. The
closed string coupling is now the same as the open string coupling
${\hat G}_s$. The axion $\hat \chi$ for the present case is independent
of the $u$. In this case $e^{\hat \phi}\to0$ in UV limit.

	In the previous section, we concluded for case I that the S-dual 
of NCOS
always corresponds to NCYM, from the open string
viewpoint. In the above, we have shown that this is also true from the
gravity (or closed string) viewpoint of D-branes. In terms of the
effective description of the closed strings discussed in footnote 9, the
above metric keeps the same form except for replacing the $\a'$ by the
 effective $\a'_{\rm eff}$. This can be checked easily as follows.
Effectively, the new closed string metric is related to the old one
given in \eqn{22} as $d {\hat s}^2_{\rm eff} = d {\hat s}^2 /
(\a'/\tilde b')^{(1 + \delta)/2}$. In other words, the $\a'$ in metric
$d {\hat s}^2$ is replaced by $\tilde b' (\a'/\tilde b')^{(1 -
\delta)/2}$ which is just the $\a'_{\rm eff}$ defined in footnote 9.
In the previous
section, we also showed that the story for NCYM is different. The S-dual
of NCYM is not necessarily an NCOS theory. We will show that
the same conclusion can be drawn
 from the gravity consideration as well in the following. 

Let us present first the gravity dual
description of the NCYM for case II discussed in the previous section.
 We list the decoupling limit collectively as    
\bea
&& \a'\to0,~~~u={r\ov \alpha'}={\rm fixed},~~g_2 = 
\left(\frac{\alpha'}{\tilde b}\right)^2,~~~cos\theta={\alpha'\over
\tilde b},\nnu\br
&& g_1 cos^2 \a = 1,~~R^4= {\rm fixed}= 4\pi g_s n {cos\a\over
cos\theta} = 4\pi G_s n
\lll{23}
\eea
As is understood, the parameters $\tilde b$ and $\tilde b'$ (introduced
later in the discussion of the S-duality of the NCYM) here are
different, in definition, from those in the previous case. The closed string 
coupling $g_s$ is related to the fixed open string coupling $G_s$ as
given in \eqn{23}.
The harmonic functions take the form;
\be
H={g_1 R^4\over u^4 \alpha'^2},~~H'=1+{R^4\over\tilde
b^2 u^4},~~H''={R^4\over\a'^2 u^4}.
\lll{24}\ee
So the metric, dilaton,  axion,  2-form $B$-field and the RR
2-form $A_2$ in eqs.\eqn{17n}
reduce to,
 \bea
&& ds^2 = \a'\left\{ {u^2\over R^2}\left[- (d\tilde
x^0)^2+ (d\tilde x^1)^2  + \bar{h}
((d\tilde x^2)^2
+ (d\tilde x^3)^2)\right] + R^2 \left[\frac{du^2}{u^2}+
d\Omega_5^2\right]
\right\} \nnu\br
&& e^{2\phi}= G_s^2 \bar{h},~~ \chi = \frac{\tilde b}{\alpha'} G_s^{-1} 
sin\a, \nnu\br
&&2\pi \a' B = \a'^2 { u^4 sin\a \over R^4}d\tilde x^0\wedge d\tilde x^1 - \a' 
{\tilde b u^4\over R^4}{\bar h} d\tilde x^2\wedge d\tilde x^3,
\nnu\br
&&A_2 = \a' \frac{\tilde b  u^4}{g_1 G_s R^4}d\tilde x^0\wedge d\tilde x^1 +
 \frac{{\tilde b}^2 u^4 sin\a}{G_s R^4} {\bar h}d\tilde x^2\wedge
d\tilde x^3,
\lll{25}\eea
where we have defined $\bar{h}={1\over 1+a^4 u^4}$ with $a^4={\tilde
b^2\over R^4}$ and $g_1$ is related to $sin\a$ as $g_1 = 1/(1 - sin^2\a)$. 

Again we note that the metric in \eqn{25} 
 has $AdS_5\times S^5$ form as $u\to0$ as expected and its form starts 
deviating from $AdS_5\times S^5$ at $u\sim R/\sqrt{b}$. From 
\eqn{25} we find that as $u\to0,~e^{2\phi}=G_s^2$ and so, the open
 string coupling is again the same as the closed string coupling at the
 IR. As $u\to\infty,~e^{2\phi}\to0$. Notice that the metric and dilaton
 are independent of the scaling factor $g_1$ but all the other fields do 
depend on $g_1$. As we can see from the above, the $01$-component of 
$B$-field depends, i.e., the  electric field depends on $g_1$. But this
field plays no role for the decoupling limit since its ratio to the
 magentic field is like $\sim \a'$, vanishing in the decoupling limit.
Notice that the axion is constant, independent of $u$. However, it blows
up in the decoupling limit unless $sin\a/\a' \to 0$ or fixed
 value. The behavior of axion is very important in determining whether
 we have good or bad S-dual theory.  This is because under
 S-duality, the closed string coupling in the dual theory is determined
by the dilaton and the axion in the original theory as $e^{\hat \phi} =
e^\phi (\chi^2 + e^{- 2 \phi})$. So if $\chi$ blows up with a fixed
 $e^\phi$ this will imply that the S-dual closed string coupling $e^{\hat
 \phi}$ always blows up in the decoupling limit. This in turn implies that
the
 open string coupling blows up since it is the same as the closed string
coupling at the IR. If this happens, we have neither a good field theory
nor a good gravity dual description. We have shown the former in the previous 
section. We will show the case for the gravity description in the
 following. So the behavior of the axion in the original theory
 determines precisely whether we have a good S-dual theory or not even
 though the original theory is nicely behaved (which is NCYM here).
The behavior of the RR 2-form potential is also consistent with this
 even though it is not directly relevant to the perturbative open string
dynamics, i.e., NCYM, in the decoupling limit.
 
	Keeping this in mind we will now discuss the S-duality of the
gravity dual description of
the NCYM discussed above. As in the previous section (Case II), we have
three cases here also
depending on the angle $\a$ or the scaling factor $g_1$: 
\begin{itemize}
\item
1) 
$g_1 = {\rm fixed} \neq 1 $, 
\item
2) $cos\a \to 0$ as $\a' \to 0$ 
\item
3) $sin\a \to 0$ as $\a' \to 0$. 
\end{itemize}

Case 3) can also be divided into three
subcases which are determined by the order parameter $\b$ defined as 
$sin\a = (\a'/\tilde b')^\b$. As studied in the previous section, cases
1), 2) and $ 0 < \b < 1$ in case 3) all give ill-behaved but ordinary YM
theories. The gravity description for these cases can be given in a
unified way as

\bea
&& d{\hat s}^2 = \frac{\tilde b\, sin\a}{G_s} \left\{ {u^2\over R^2}
\left[ -(d\tilde
x^0)^2+ (d\tilde x^1)^2 + \bar{h}
( (d\tilde x^2)^2
+ (d\tilde x^3)^2)\right] + R^2 \left[\frac{du^2}{u^2}+
d\Omega_5^2\right]
\right\} \nnu\br
&& e^{\hat \phi}=\frac{sin^2\a}{G_s} \left(\frac{\tilde b}{\a'}\right)^2
 \bar{h}^{1/2},~~ \hat\chi = - \frac{G_s}{sin\a}\frac{\a'}{\tilde b} 
 \lll{26}
\eea
where all the quantities are defined as in \eqn{18n} and again 
$2\pi \alpha' \hat B = A_2$ and $\hat A_2 = - 2\pi\a'B$ with $B$ and
$A_2$ given in \eqn{25}.
 For each of these three cases, 
we can check that the
closed string coupling always blows up as $\a' \to 0$ precisely because 
$sin\a /\a' \to \infty$ as $\a' \to 0$ as discussed above. So the
gravity description breaks down. At the IR, the closed string coupling 
$e^{\hat \phi}$ is just the open string coupling given in the previous 
section which also blows up. So the underlying theory is not good
even though the original NCYM theory appears fine. 

	For other two sub-cases in case 3), i.e. for $\beta\ge 1$, we have
good gravity descriptions as 
expected. They can also be given in a unified way as
\bea
&& d{\hat s}^2 = \a' \left(\hat G_s /G_s\right)^{1/2} \bar
h'^{-1/2}\times\nnu\br 
&&\left\{ {u^2\over R^2}\left[-d\tilde
x_0^2+ d\tilde x_1^2 + \bar{h}
(d\tilde x_2^2
+ d\tilde x_3^2)\right] + R^2 \left[\frac{du^2}{u^2}+
d\Omega_5^2\right]
\right\} \nnu\br
&& e^{\hat \phi}= \hat G_s \frac{\bar h^{1/2}}{\bar h'},~~ 
\hat\chi = - \hat G_s^{ -1}  \frac{\tilde b}{\tilde b'}
\left(\frac{\a'}{\tilde b'}\right)^{\b - 1} \bar h' 
 \lll{27}
\eea
where we have defined $\bar h' = 1/(1 + a'^4 u^4)$ with $a'^4 = \tilde
b^2 /(R^4 \hat G_s G_s)$ and $\bar h$ is the
same as that given for NCYM.  
We have $\hat G_s G_s = [1 + (\tilde b/\tilde b')^2]$, for
$\b = 1$ and $\hat G_s G_s = 1$, for $\b > 1$. Again the metric reduces
to $AdS_5\times S^5$ as expected and its form starts deviating from 
$AdS_5\times S^5$ at $u \sim R/\sqrt{\tilde b}$. The open string
coupling $\hat G_s$ is the same as the closed string coupling at the IR.
Unlike in the original gravity description, the closed string coupling 
blows up at the UV.

\section{A Few Remarks}

	In the previous sections, we have studied the relationship
between NCOS and NCYM using the  BPS ((F, D1),
D3) bound state configuration in two versions related by S-duality. For
concreteness,
we choose the asymptotic string-frame metric in Version A as 
$\eta_{MN} = (-, +, \cdots, +)$, where $M, N = 0, 1, \cdots, 9$
with respect to the unscaled coordinates.
 Because of this choice, 
the string-frame metric in
the S-dual version (i.e., Version B) is not
$\eta_{MN}$ asymptotically but to $g_s^{-1} \eta_{MN}$ with respect to
the unscaled coordinates. 

In the previous section,
 we have given the gravity dual descriptions of NCOS (or NCYM) and 
its S-dual in Case I (or Case II). 
In Case I,
we always have $r \sim \sqrt{\a'} u$ while in Case II we have
 $r = \alpha' u$.
The scalings for $r$ are different in the two cases\footnote{
We don't have 
this difference if we study the NCYM in Version B (rather than in
 Version A) in Case II. However, the above choice helps us to understand 
 things better and that is why we prefer to present in this way.}.
How can we understand this difference? Another puzzle is: when we
discuss the condition in Version B for NCOS in Case II  as the S-dual of 
NCYM in Version A, we have (see eqn\eqn{29n}) 
\be
\frac{\hat b_1} {\hat b_2} \sim \frac{1}{\a'^{1 + \b}},\lll{28}
\ee
 with $\b \geq 1$. While in Case I 
 for NCOS in Version A, we have  
\be
\frac{b_1}{b_2} \sim \frac{g_1 sin\a} {g_2
tan\theta} \sim \frac{1}{\a'^{1 - \delta + \delta'}}, \lll{29}
\ee
 where the last equality is
obtained using eqs. \eqn{20nn}-\eqn{21n} and we also use $sin\theta \sim
\a'^{\delta'}$ with $\delta' \geq 0$. These two criteria should be the
same but from their appearances, it does not seem to be the case. Let us
understand these two puzzles at the same time.

For convenience, let us collect some scalings here: For case I,
we have
\be 
r \sim \sqrt{\a'} u,~~ g_1 \sim \a'^\delta,~~ g_2 \sim \a',~~
sin\theta \sim \a'^{\delta'}, g_s^{-1} \sim \a'^{(1 - \delta)/2}
\lll{30}\ee
where $\delta < 1$ and $\delta' \geq 0$. While for Case II, we have
\be 
r = \a' u,~~g_1\sim 1,~~ g_2 \sim \a'^2,~~g_s \sim \a',~~\sin\a \sim \a'^\b,
\lll{31}\ee
where $\b \geq 1$.

	The string constant $\alpha'$ is a property of the string,
independent of the background in which it moves. However, the scaling
behavior for the radial coordinate $r$ in a gravity dual description 
in terms of $\a'$  does depend on the asymptotic metric.
We also know that if $\alpha'$ is the string constant
 defined through $1/(4\pi \alpha') \int \partial X^M \partial
X^N \eta_{MN}$ and $r$ is one of the coordinates $X^M$, then for NCYM,
we should have $r = \alpha' u$ while for NCOS, we should have 
$r =\sqrt{\a'}\, u$ \cite{luroytwo,gopmms}. So for the NCOS in Case I (in 
Version A) 
and for NCYM in Case II (also in version  A), we do have the correct
scaling
behavior. They can be further understood in the following way. For the NCOS,
the string lies along $x^1$-direction and the $\a'$ is defined with
respect to the unscaled $x^1$. With respect to the scaled $\tilde x^1$,
we have $1/(4\pi \alpha') \int \partial x^1 \partial x^1 = 1/(4 \pi \a')
\int \partial {\tilde x}^1 \partial{\tilde x}^1 g_1$. So the effective
string constant $\a'$ is $\a'_{\rm eff} = \a' g_1^{ - 1}$. 
In order to have the same
effective $\a'_{\rm eff}$ for the radial coordinate $r$, 
we must rescale $r = \sqrt{g_1} {\tilde r}$. 
We expect that $\tilde r \sim \sqrt{\a'_{\rm
eff}} \,u$. It is easy to check, using the above relations, that this is 
indeed true. For the NCYM, $g_1 \sim 1$ and we always have $r \sim \a'
\,u$. So for either the NCOS or the NCYM, the scaling behavior
remains the same whether we use $\a'$ or $\a'_{\rm eff}$.
With the above in mind, we  now try to understand $r \sim
\sqrt{\a'} \,u$ for NCYM in Case I and $r = \a'\, u$ for NCOS in Case
II, both of them in Version B. 

The asymptotic metric for $g_{11}$ with respect to the
scaled coordinate in Version B is $g_s^{-1} g_1$. So we have $1/(4\pi \a') \int 
\partial x^1 \partial x^1 g_s^{-1} = 
g_s^{-1} g_1/(4\pi \a') \int \partial {\tilde x}^1 \partial {\tilde
x}^1$. So the effective string constant $\a'_{\rm eff} = \a' g_s /g_1$.
In order to have the same effective constant for coordinate $r$, we 
must again rescale the $r$ as 
$\tilde r = r/\sqrt{g_1}$.

For the NCYM in Case I, we apply the above two relations and we have
$\a'_{\rm eff} \sim \a'^{(1 - \delta)/2}$ where eq. \eqn{30} has been
used, which is also consistent with that given in footnote 9.
This gives the rescaled $\tilde r \sim \sqrt{\a'} u/\sqrt{g_1} 
\sim \a'^{(1 - \delta)/2} u \sim \a'_{\rm eff} u$, the expected relation. 
While for the NCOS in Case II, we have $\a'_{\rm eff} \sim \alpha'^2$. This
gives $\tilde r = \a' u \sim \sqrt{\a'_{\rm eff}}\, u$, 
again the correct relation.

	Even though we discuss the scaling of $r$ for both Case I and
Case II at the same time, so far we have not brought any connection
between them
yet. Even then we expect that the conclusions drawn in each case should be
the same. But the ratio between the electric and magnetic fields,
determining NCOS, looks
different in the two cases. How can we resolve this puzzle? 
Note that $\a'$ used in defining the ratio in Case I can be taken as
the effective string constant with respect to the unscaled coordinates 
for the NCOS in this case while the $\a'$ used in defining the same ratio in
Case II can be taken as the effective string constant with respect to the
unscaled coordinates for the NCYM in that case. In other words, the
$\a'$ is taken as the effective string constant for different theories
in the two cases. This is the source of
difference. We expect that if the same $\a'$ is used as an effective
string coupling for either NCYM or NCOS in both Case I and Case II, the
ratio should scale the same way. This suggests that if expressed in
terms of the effective
string constant $\a'_{\rm eff} \sim \a'^{(1 - \delta)/2}$ for the NCYM in
Case I, the ratio should scale in the same way as that in Case II. Let us
check 
if this is true. With $\a'_{\rm eff} \sim \a'^{(1 - \delta)/2}$, we
have the ratio in Case I, from \eqn{29}, as
\be
\frac{b_1}{b_2} \sim \frac{1}{\a_{\rm eff}^{'\,2 + 2 \delta'/(1 -
\delta)}},\lll{32}
\ee
with $\delta < 1, \delta' \geq 0$. Since $\delta < 1$, so $\a'_{\rm eff}
\to 0$ as $\a' \to 0$. This ratio has the same behavior as that in Case
II if we identify the $\a'$
there as our present effective $\a'_{\rm eff}$  and
$\b$ there as  $\b = 1 + 2 \delta'/(1 -
\delta)$. Note that $\b \geq 1$  is consistent with
 $1 + 2\delta'/(1 - \delta) \geq 1$, since $\delta' \geq 0$ and $\delta <
1$. In particular, $\b = 1$ corresponds to $\delta' = 0$, both of which
give their respective NCOS's with noncommutativity in both space-space
and space-time directions. 

To identify the two cases completely, we need
to set $g_1 \sim 1$ for NCYM in Version A of Case II the same as 
 $g_s g_1 \sim \a'^{(1 + \delta)/2}$ for NCYM in Version B of Case I.
 This implies that $\delta = - 1$. Now $\a'_{\rm eff} = \a'$
as expected. This is consistent with the fact that the NCYM in Version A
of Case II has fixed metric while the NCYM in Version B of Case I has
fixed metric only for $\delta = - 1$.
One can check that the scalings of the relevant quantities for
NCOS in Version A of Case I are the same as those for the NCOS in
Version B of Case II (similarly
for the NCYM's in both cases). For example, $g_2 \sim \a'$ for the NCOS
in Version A of Case I while the equivalent one for the NCOS in Version
B of Case II is $g_s^{-1}
g_2 \sim \a'^{-1} \a'^2 \sim \a'$ as expected.

Finally in this section, we discuss the possible quantization for the
open string coupling constant $G_s$ and $\hat G_s$ in the decoupling
limit. One can use the definitions for $cos\theta$ and $cos\a$ in
\eqn{17nb}
to express the open string coupling $G_s$ as
\be
G_s = g_s \frac{cos\a}{cos\theta} = g_s \frac{q^2 + n^2}{n \sqrt{(p
g_s)^2 + q^2 + n^2}},\lll{33}
\ee
and the S-dual $\hat G_s$ in \eqn{23n} as
\be 
\hat G_s = g_s^{-1} \left(1 - sin^2\theta\,cos^2\a\right)^{1/2}
\left(1 + \frac{tan^2\a}{cos^2\theta}\right)^{1/2} = g_s^{-1} 
\frac {\sqrt{(p g_s)^2 + n^2}}{\sqrt{(p g_s)^2 + q^2 + n^2}} \left(1 + 
\frac{(p g_s)^2}{n^2}\right)^{1/2}. \lll{34}
\ee
The integral charge $n$ for D3
branes is always fixed.  In the case $p g_s \gg \sqrt{q^2 + n^2}$, we have
\be 
G_s = \frac{q^2 + n^2}{n p},~~\hat G_s = \frac{p}{n}.\lll{35}
\ee
In other words, both $G_s$ and $\hat G_s$ are independent of the closed
string coupling and quantized in the above
limit. The question is now: Can both $G_s$ and $\hat G_s$ be finite and
be related to the decoupling limits discussed in the previous sections?

	Let us discuss the Case I first. To the NCOS, the integer $p$ is 
related to the electric field and the integer $q$ is related to the
magnetic field while to its S-dual, i.e., NCYM, these relations are
exchanged, i.e., $q$ to the electric field and $p$ to the magnetic field.
The decoupling limit for this case says that $cos \a \to 0,~ cos\theta
\sim 1$ and $g_s \to \infty$. They imply that $ q\sim n = {\rm fixed}$ 
and $p g_s \gg \sqrt{n^2 + q^2}$, the right condition to validate the 
quantizations of both $G_s$ and $\hat G_s$ while at the same time, $p$
can be finite. So the decoupling limit in Case I for NCOS/NCYM gives 
finite and quantized open string coupling $G_s$ and its S-dual $\hat
G_s$. Also if $q = 0$, i.e., one of the field vanishes, we have 
$\hat G_s G_s = 1$, the expected relation.

	When NCYM and NCOS are S-dual to each other in Case II,
 we have $cos \theta \to 0,~ cos\a \sim 1$ and $g_s \to 0$.
 This decoupling limit does not give the condition needed
for $G_s$ and $\hat G_s$ to be independent of the closed string
coupling. Therefore both $G_s$ and $\hat G_s$ cannot be expressed purely
 in terms of the charges $p, q$ and $n$. However, for the subcase 2 in
 Case II for which the S-dual of the
 NCYM  is a singular ordinary Yang-Mills, we have $cos\theta \to 0,
cos\a \to 0$ which implies that $pg_s \gg \sqrt{q^2 + n^2}$ and $
q \gg n$. In this case, we also have both $G_s$ and $\hat G_s$ to be
 independent of the closed string coupling $g_s$ and they are given as
\be
G_s = \frac{q^2}{pn},~~\hat G_s = \frac{p}{n},
\ee
which gives $G_s \hat G_s = q^2/n^2 \gg 1$, as expected.

	Let us now try to understand when NCOS and NCYM are S-dual 
to each other, why in Case I the
open string couplings are independent of the closed string coupling
while in Case
II they are not? Our possible explanation is as follows: In case I, we
choose the fundamental string-frame in Version A and we end up with NCOS
as a fundamental open string because of the critical electric field. In
other words, the tension for the D3 branes in the decoupling limit for
NCOS must be independent of the closed string coupling. Therefore, the 
corresponding gauge coupling is also independent of the closed string
coupling, which in turn implies that  the open string coupling is independent
of the closed string coupling. This can also be 
understood for its S-dual, i.e., NCYM, in Version B of Case I. Given
that the Version A is expressed in the fundamental string frame, Version
B is therefore expressed actually in D-string frame. So for a D-string
in Version B, its tension is independent of the closed string coupling
\footnote{The tension for a brane in its own frame is always independent
of the closed string coupling.}. Since the S-dual of NCOS is NCYM with a 
rank 2 $B$-field, this NCYM is equivalent to an ordinary Yang-Mills 
in $1 + 1$ dimensions with $U (\infty)$ gauge group whose gauge coupling 
is determined by the D-string tension \cite{luroyone}. But this D-string 
tension in Version B is independent of the closed string coupling. So
the open string coupling must be also independent of the closed string 
coupling.

	With the above, it is not difficult to understand why in Case
II, the open string couplings are dependent on the closed string
coupling since now the NCOS as a fundamental string moves in a D-string
frame while NCYM is defined in a fundamental string frame.

\bigskip
{\it Note added}: After the submission of this paper, there were two
related papers which appeared in the net \cite{russj,kawt}. 

\bigskip
\centerline{\bf ACKNOWLEDGMENTS}

We thank our referee for his comments which help us to improve the manuscript.
JXLU acknowledges the support of U. S. Department of Energy.

\end{document}